\documentclass[aps,prev,twocolumn,preprintnumbers,floatfix,nofootinbib]{revtex4-1}
\usepackage[scientific-notation=true]{siunitx}
\usepackage{float, verbatim, amsmath, amssymb, fullpage, paralist, cancel, listings, subfig, enumitem, siunitx, graphicx, xcolor, tikz, booktabs, tcolorbox, textcomp, url, parse lines, fontawesome5, marginnote, smartdiagram, forest, pifont, pgfcalendar, pgfgantt, fix-cm, ulem, soul, multirow, xfrac, colortbl, xspace, fontenc, helvet, mathptmx,ragged2e,csquotes} 
\definecolor{smcolor}{rgb}{0.8,0.5,0.5}
\definecolor{bsmcolor}{rgb}{0.4,0.6,0.8}
\definecolor{pvcolor}{rgb}{0.8,0.8,0.8}
\definecolor{cobalt}{rgb}{0.0, 0.28, 0.67}
\definecolor{darkelectricblue}{rgb}{0.33, 0.41, 0.47}
\definecolor{darkpowderblue}{rgb}{0.0, 0.2, 0.6}
\definecolor{darktangerine}{rgb}{1.0, 0.66, 0.07}
\definecolor{pastelviolet}{rgb}{0.8, 0.6, 0.79}
\definecolor{black}   {RGB}{0.0, 0.13, 0.28}
\definecolor{dukeblue}    {rgb}{0.0, 0.0, 0.61}
\definecolor{oxfordblue}{rgb}{0.0, 0.13, 0.28}
\definecolor{navyblue}{rgb}{0.0, 0.0, 0.5}
\definecolor{linkred}{RGB}{165,0,33}

\usesmartdiagramlibrary{additions}
\usepackage{hyperref}
\usepackage{cleveref}[capital]
\usepackage{textcase}
\hypersetup{
        linktocpage=true,
        setpagesize=true,
	colorlinks= true,
	linkcolor=cobalt!50!black,
        menucolor=cobalt!50!black,
	filecolor=cobalt!50!black,      
	urlcolor=cobalt!50!black,
        citecolor=cobalt!50!black,
        citebordercolor=cobalt!50!black,
	pdftitle={Light Dark Scalar},
        pdfsubject={Latex},
        pdfauthor={Yoxara Villamizar},
        pdfkeywords={BSM Phenomenology, SM, Collider, DM}
}
\graphicspath{{Figures/}} 
\newlist{firstitemize}{itemize}{1}
\setlist[firstitemize,1]{label=\color{cobalt!70!black} \ding{111}, nosep, leftmargin=*, before=\normalcolor}
\newlist{seconditemize}{itemize}{1}
\setlist[seconditemize,1]{label=\color{cobalt!70!black}\ding{226}\normalcolor, nosep, leftmargin=*} 
\newlist{thirditemize}{itemize}{1}
\setlist[thirditemize,1]{label=\color{cobalt!70!black} \ding{118}, nosep, leftmargin=*, before=\normalcolor}

\renewcommand{\bar}{\overline}

\usepackage{tikz}
\usetikzlibrary{arrows.meta,decorations,calc,shadows.blur,decorations.pathreplacing,decorations.pathmorphing,decorations.markings,shapes, arrows, positioning,bending,chains,matrix, patterns,mindmap,trees}
\tikzset{
>=triangle 45,
scalar/.style={decorate, dashed, draw=cobalt!50!black, thick, line width=0.8pt}, 
photon/.style={decorate, line width=0.85pt, draw=black, decoration={snake, segment length=5, amplitude=3pt, post length=0.5mm, pre length=0.5mm}}, 
particle/.style={draw=black, postaction={decorate}},
fermion/.style={draw=black, line width=0.6pt, postaction={decorate}, decoration={markings,mark=at position .5 with {\arrow[draw=black,scale=1.2,>=stealth]{>}}}},
antifermion/.style={draw=black, line width=0.6pt, postaction={decorate},decoration={markings,mark=at position .5 with {\arrow[draw=black,scale=1.2,>=stealth]{<}}}},
}

\usepackage{graphicx}
\usepackage{amsmath, amsthm, amssymb}
\usepackage{amsfonts}
\usepackage{subfig}
\usepackage{xspace}
\usepackage[dvipsnames]{xcolor}
\usepackage{paralist}
\usepackage{multirow}
\usepackage{paralist}
\usepackage{graphicx}
\usepackage{bm}
\usepackage{times}
\usepackage{slashed}
\usepackage{color}
\usepackage{epsfig}\usepackage{dcolumn}
\usepackage{float}

\usepackage{makecell}

\usepackage{color}
\def\br{\begin{eqnarray}}
\def\er{\end{eqnarray}}
\def\be{\begin{equation}}
\def\ee{\end{equation}}

\begin{document}

\title{Can the FCC-hh prove the B-L gauge symmetry?}

\author{Farinaldo S. Queiroz$^{1,2,3,4}$}
\author{J. Zamora-Saa$^{2,5}$}
\author{Ricardo C. Silva$^{1,3,6}$}
\author{Y.M. Oviedo-Torres$^{1,2,}$}

\email{Corresponding author: mauricio.nitti@gmail.com}

\affiliation{
$^1$International Institute of Physics, Universidade Federal do Rio Grande do Norte, Campus Universit\'ario, Lagoa Nova, Natal-RN 59078-970, Brazil, \\
$^2$Millennium Institute for Subatomic Physics at High-Energy Frontier (SAPHIR), Fernandez Concha 700, Santiago, Chile, \\
$^3$Departamento de F\'isica Teorica e Experimental, Universidade Federal do Rio Grande do Norte, 59078-970, Natal, Rio Grande do Norte, Brazil, \\
$^4$Departamento de F\'isica, Facultad de Ciencias, Universidad de La Serena, Avenida Cisternas 1200, La Serena, Chile,  \\
$^5$Center for Theoretical and Experimental Particle Physics - CTEPP, Facultad de Ciencias Exactas, Universidad Andres Bello, Fernandez Concha 700, Santiago, Chile, \\
$^6$Laboratoire de physique nucléaire et des hautes énergies (LPNHE), Sorbonne Université, Paris, France.
}

\begin{abstract}

\noindent We present a phenomenological study of the discovery potential at the FCC-hh for a new heavy neutral vector boson, $Z^{\prime}$, predicted by the $U(1)_{B-L}$ gauge symmetry. Focusing on the parameter space currently not excluded by Large Hadron Collider data, we analyze the dilepton production channel $p p \rightarrow Z^{\prime} \rightarrow l^{+} l^{-}$ ($l^{\pm} = e^{\pm}, \mu^{\pm}$) at a center-of-mass energy of $\sqrt{s} = 100$ TeV. Full Monte Carlo simulations was performed for different ($M_{Z'}$, $g_{B-L}$) BSM scenarios and relevant Standard Model backgrounds (including irreducible Drell-Yan, diboson, single top-quark and top-quark pair productions) identifying optimal kinematic and angular selection cuts to guide future searches for this type resonance. We estimate the FCC-hh reach for an integrated luminosity of $\mathcal{L}_{int}$ = 3~$ab^{-1}$. Our results demonstrate that the FCC-hh can exclude $Z^{\prime}$ masses  up to $\sim 40$ TeV with 95\% C.L. for couplings of $g_{B-L} \sim 1$, and up to $\sim 15$ TeV for $g_{B-L} \sim 0.1$. We find the kinematic and angular cuts that optimize the signal over background ratio and achieve a $5\sigma$ signal up to Z' masses of $\sim 30$ TeV. These findings highlight the FCC-hh potential to uncover new physics signals in the high-mass regime.
\end{abstract} 
    
\maketitle

\section{Introduction}
\label{introduction}

The Future Circular Collider in its hadron-hadron mode (FCC-hh) represents a transformative step forward in high energy physics, designed to operate at a center-of-mass energy of 100~TeV with an integrated luminosity of up to 30~ab$^{-1}$~\cite{FCC:2018vvp,ParticleDataGroup:2024cfk}, the FCC-hh is poised to explore new regimes of energy and precision, far beyond the current reach of the Large Hadron Collider (LHC). The development of the Standard Model (SM) has been intimately shaped by successive advancements in collider technology. From the discovery of partonic structure at SLAC to the observation of $W/Z$ bosons at SPS \cite{Shiltsev_2012, Bloom:1969kc, UA1:1983crd, UA1:1983mne}, a series of achievements, such as the discovery of the top quark at the Tevatron, precision physics at LEP, and the discovery of the Higgs boson at the LHC, have established the internal consistency of the SM with remarkable precision \cite{PhysRevLett.74.2626, ALEPH:2006bhb, ATLAS:2012yve}. These milestones illustrate that progress in accelerator technology has been instrumental in advancing particle physics, and the FCC continues to this legacy by pushing the energy frontier to a new scale. 

The FCC is proposed to be housed in a new circular tunnel with a circumference of 90 to 100~km, significantly expanding the discovery potential for new heavy states well beyond those probed by the current LHC, and its high luminosity could even provide sensitivity to rare processes with small cross-sections. The FCC program is planned in stages, beginning with an electron-positron collider (FCC-ee) in the 2040s, followed by the installation of the FCC-hh after a decade-long transition. This staged strategy aims to deliver a comprehensive 50 year experimental program until the end of the 21st century. With its extensive discovery reach, the FCC program will enable precise measurements of the Higgs sector, such as the Higgs self-coupling \cite{stapf_2025_c94hf-bqc82,lipeles_2025_98taa-y9506}, while probing the top-quark and electroweak sectors with unprecedented precision \cite{Mangano:2016jyj}. Furthermore, it will allow for the exploration of new bosonic and fermionic resonances, providing direct probes into various new physics scenarios \cite{Golling:2016gvc, Blondel:2025kbl}. The FCC project can even contribute to the production and analysis of new particles that could potentially be discovered in the near future at the High-Luminosity LHC (HL-LHC) \cite{CidVidal:2018eel}. \newline

Despite its tremendous success, the SM still fails to explain open problems such as the origins of neutrino masses, the matter-antimatter asymmetry~\cite{Cvetic:2021itw,Cvetic:2015naa,CarcamoHernandez:2022fvl,deJesus:2023lvn,Cogollo:2023twe}, the elusive nature of dark matter~\cite{Bertone:2004pz,Alves:2013tqa,Bertone:2016nfn,Arcadi:2017kky,Arcadi:2024ukq}, among others. The extension of the SM that promotes baryon and lepton numbers to be gauge symmetries known as \(U(1)_{B-L}\) ~\cite{Davidson:1978pm,Mohapatra:1979ia, Ma:1998dx,Mohapatra:1986bd} has been the target of numerous collider studies \cite{Emam:2007dy,Basso:2008iv,FileviezPerez:2009hdc,Li:2010rb,Arcadi:2013qia,Abdelalim:2014cxa,Alves:2015mua,Klasen:2016qux,Lindner:2016lpp,Lindner:2016bgg,VanDong:2018yae,Lindner:2018kjo,Camargo:2018uzw, Camargo:2018klg,Lindner:2020kko}. In this framework, the introduction of three right-handed neutrinos ensures anomaly cancellation, and a new neutral vector boson, the \(Z^\prime\), naturally emerges. A key feature of the \(U(1)_{B-L}\) model is that the \(Z^\prime\) boson couples to all SM fermions proportionally to the B-L quantum numbers, which is 1/3 for quarks and unit for leptons and gauge coupling $g_{B-L}$. This implies that the phenomenology of the \(Z^\prime\) is largely determined by its mass and overall coupling constant, making it an ideal target for searches at colliders. Although current LHC searches have probed \(Z^\prime\) masses up to around 5--6~TeV in clean dilepton channels~\cite{Cox:2017eme, CMS:2023ooo}, the FCC-hh's superior energy and luminosity will enable sensitivity to much heavier \(Z^\prime\) bosons and to scenarios with smaller coupling strengths.

We simulate the SM background and signal events for various \(Z^\prime\) masses and coupling strengths, and analyze the FCC-hh sensitivity in the dilepton channel, comparing these projections with current experimental limits. To this end, we perform a detailed phenomenological analysis employing optimized selection cuts designed to enhance signal-to-background discrimination. These cuts are tuned to improve both exclusion and discovery reach for given benchmark scenarios and integrated luminosities. This approach allows us to derive more realistic and robust projections for the reach of the FCC-hh in probing the \(B-L\) gauge sector through high-mass dilepton signatures. 

This paper is organized as follows: in Section~\ref{fcchh}, we present an overview of the Future Circular Collider and its experimental program. In Section~\ref{blmodel}, we briefly review the relevant theoretical aspects of the \(U(1)_{B-L}\) model. In Section~\ref{results_and_discussions} we present the results and discussions where we discuss in detail the simulations of $Z^\prime$ production in the dileptonic channel as well as the different background sources, and the strategy followed to obtain the required luminosity to exclude or discover $Z^\prime$ bosons in the FCC-hh for the first years of operation. Finally, we conclude in Section~\ref{conclusions}. \newline

\section{The Future Circular Collider: $\text{FCC-hh}$}
\label{fcchh}

The FCC-hh is a proposed next-generation proton-proton collider envisioned by CERN as the natural continuation of the LHC program. It is designed to operate at an unprecedented center-of-mass energy of 100~TeV, with a target integrated luminosity of up to 30~$ab^{-1}$~\cite{FCC:2018vvp}, including sensitivity for heavy resonances with masses up to 50 TeV \cite{Helsens:2019bfw}. This configuration will enable exploration of physics far beyond the capabilities of the LHC and its HL-LHC upgrade\footnote{Although the total expected luminosity for the FCC-hh for 30 years of operation is 30~$ab^{-1}$, in this work we will demonstrate that with the first years of operation, a luminosity of 3~$ab^{-1}$ is enough to probe masses up to 30 TeV with up to 5$\sigma$ of statistical significance.} \cite{FCC:2025lpp}. 

The FCC-hh will be housed in a circular tunnel with a circumference of approximately 90--100~km, significantly larger than the 27~km ring of the LHC. The collider infrastructure will include large underground experimental caverns for detectors, with vertical shafts to facilitate construction and access. The reference detector design anticipates a cylindrical geometry wich has a diameter of 20~m and a length of 50~m, comparable to the dimensions of the ATLAS detector at the LHC. Besides that it has a central and forward detector with pseudorapidity coverage up to $|\eta| \sim 6$, optimized for the identification of extremely boosted final states \cite{FCC:2025uan}. 

The FCC project is structured in stages. Initially, an electron-positron collider (FCC-ee) will be installed in the tunnel, serving as a Higgs, Z, and top factory. The FCC-ee is expected to begin operation in the second half of the 2040s and run for approximately 15 years. Following its decommissioning, the FCC-hh will be installed in the same tunnel, with a projected operational lifetime of 25 years. This staged approach enables a comprehensive physics program extending over five decades, analogous in vision to the LEP-LHC sequence \cite{FCC:2025jtd}. 

The physics potential of the FCC-hh is immense. It will extend the direct discovery mass reach to several tens of TeV, enabling the production of new particles that may have been indirectly hinted at through precision measurements performed in the preceding electron-positron phase. In addition to its discovery potential, the collider will enable precise determination of the Higgs boson self-coupling and offer a detailed exploration of electroweak symmetry breaking dynamics at the TeV scale, providing critical insights into the nature of the electroweak phase transition. Moreover, the FCC-hh will be able to probe Weakly Interacting Massive Particles (WIMPs) as thermal dark matter candidates with unprecedented sensitivity, leading either to their discovery or to stringent exclusion limits \cite{FCC:2018byv}. 

In this context, the FCC-hh provides an ideal environment for searches for new heavy gauge bosons such as the $Z^\prime$ predicted in $U(1)_{B-L}$ extensions of the SM. Its center-of-mass energy and luminosity open the possibility of probing $Z^\prime$ masses well above the exclusion limits currently set by LHC data, particularly in dilepton final states. 
\section{The Minimal $B-L$ Model}
\label{blmodel}

In the SM, both baryon number and lepton number are accidental global symmetries, not imposed by the gauge structure but emerging from the particle content and renormalizability. A natural and well-motivated extension of the SM consists of promoting these quantum numbers to local gauge symmetries, however only the combination \( B-L \) is free of gauge anomalies. Remarkably, the cancellation of all triangle anomalies, including mixed gravitational ones, is automatically achieved upon the inclusion of three right-handed neutrinos with charge \(-1\) under the \( U(1)_{B-L} \) symmetry \cite{Arcadi:2023lwc}. As a consequence, this extension not only remains theoretically consistent but also naturally accommodates a Type-I Seesaw Mechanism for neutrino mass generation, thereby addressing one of the key shortcomings of the SM.

The gauging of \( B-L \) leads to the existence of a new neutral vector boson, the \( Z^\prime \), which can be searched for at current and future high-energy colliders~\cite{Pisano:1992bxx,Alves:2023vig,Robinett:1982gw,ATLAS:2019erb,RamirezBarreto:2010vji,Cogollo:2020afo}. In the minimal \( U(1)_{B-L} \) model, the SM gauge group is extended to \( SU(3)_C \times SU(2)_L \times U(1)_Y \times U(1)_{B-L} \), and the mass of the \( Z^\prime \) boson can arise either through the Stueckelberg mechanism~\cite{Ruegg:2003ps} or via the spontaneous breaking of the \( B-L \) symmetry by a scalar singlet field, which also generates Majorana mass terms for the right-handed neutrinos in an elegant way. 
At the FCC-hh, the $Z^\prime$ boson can be produced via the Drell–Yan mechanism in proton-proton collisions, we will consider the process $pp \rightarrow X \rightarrow \ell^+ \ell^-$, where $X$ can be the photon, $Z$ boson, or the new $Z^\prime$. The clean signature of high-mass dilepton final states offers a powerful probe of this new gauge interaction at the multi-TeV scale, with the enhanced energy and luminosity of the FCC-hh, we aim to project the discovery and exclusion potential for the $Z^\prime$ boson in the minimal $B-L$ model. The relevant Lagrangian terms governing the $Z^\prime$ interactions with SM fermions are, 

%

\begin{equation*}
    \mathcal{L}\supset -\frac{1}{4}{F}^{\prime}_{\mu\nu}{F}^{'\mu\nu} + {g}_{B-L}\sum_{i=1,...,6} {Q}_{X{q}_{i}} \overline{q_{i}} {\gamma}^{\mu} q_{i} {Z}^{\prime}_{\mu}
\end{equation*}

\begin{align}
 +{g}_{B-L}\sum_{\ell=e,\mu,\tau} {Q}_{X\ell} \overline{\ell} {\gamma}^{\mu} \ell {Z}^{\prime}_{\mu}  + \, g_{B-L}\,\sum_{i=1,...,3} {Q}_{X{N}_{i}} \bar{N_i} \gamma_\mu \gamma_5 N_i {Z}^{\prime}_{\mu}
    \label{relevant_lagrangian}    
\end{align}

\noindent where $F^{\prime}_{\mu\nu}$ is the new strength tensor, $Q_{Xf}$ are the B-L charges of each fermion field $f$, $g_{B-L}$ is the coupling constant of the new $B-L$ symmetry group, ${Z}^{\prime}_{\mu}$ is the new neutral vector field, and ${N}_{i}$ are the three right-handed neutrinos\footnote{In this work we assume the right-handed neutrinos ${N}_{i}$ to be sufficiently heavier than the new ${Z}^{\prime}$ vector boson, that is ${M}_{{N}_{i}} \gg {M}_{{Z}^{\prime}}/2$. This means that right-handed neutrinos do not contribute to the ${Z}^{\prime}$ invisible decay.}. In TABLE \ref{table1}, we present the content of matter particles and their respective charges under each symmetry. 

\begin{table}[h]
\begin{tabular}{|c|c|c|c|c|c|}
\hline
Field & $SU(3)_{C}$ & $SU(2)_{L}$ & $U(1)_{Y}$ & $U(1)_{B-L}$  \\ \hline
 ${\ell}_{iL}$ & 1 & 2 & -1/2 & -1  \\ \hline
 ${\ell}_{iR}$ & 1 & 1 & -1 & -1  \\ \hline
 ${N}_{iR}$ & 1 & 1 & 0 & -1 \\ \hline
 ${q}_{iL}$ & 3 & 2 & 1/6 & 1/3  \\ \hline
 ${q}^{u}_{iR}$ & 3 & 1 & 2/3 & 1/3  \\ \hline
 ${q}^{d}_{iR}$ & 3 & 1 & -1/3 & 1/3  \\ \hline
\end{tabular}
\caption{Particles and their respective charges under the $SU(3)_{C} \times SU(2)_{L} \times U(1)_{Y} \times U(1)_{B-L}$ group.}
\label{table1}
\end{table}

In the following section we present the details and analysis of results where we estimate the sensitivity of the FCC-hh to different scenarios ($M_{Z^{\prime}}$, $g_{B-L}$) of the B-L model. 

\section{Results and Discussions}
\label{results_and_discussions}

\begin{figure}[!ht]
    \centering
    \includegraphics[scale=0.45]{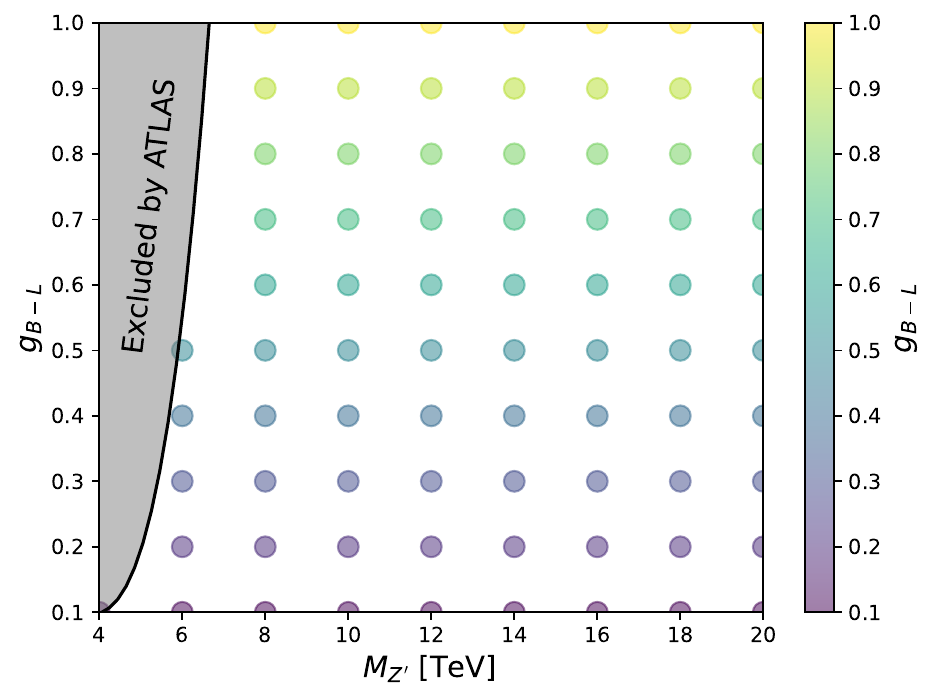}
    \caption{Different scenarios in the ($M_{Z'}$, $g_{B-L}$) plane considered for the sensitivity analysis at the FCC-hh. The gray region represent the parameter space from the $B-L$ model already excluded by current LHC searches \cite{Queiroz:2024ipo}.}
\label{cross_section_benchmarks_a}
\end{figure}

\noindent Based on the exclusion limits for a new $Z^{\prime}$ vector boson from the $B-L$ model obtained in \cite{Queiroz:2024ipo}, this analysis focuses on the parameter space not yet excluded by current LHC results \cite{ATLAS:2019erb}. This region is shown in Fig. \ref{cross_section_benchmarks_a} in the ($M_{Z^{\prime}}$, $g_{B-L}$) plane, where each colored dot represents a possible benchmark scenario to be explored. 

\noindent For each one of these not excluded scenarios, we will estimate the discovery potential of the FCC-hh ($\sqrt{s} = 100~\mathrm{TeV}$) for a new $Z^{\prime}$ vector boson predicted by the $B-L$ model, and we will show that it is possible to achieve levels of statistical significance sufficient to exclude scenarios with a 95\% confidence level or, in a discovery scenario, to achieve 5$\sigma$ in most scenarios ($M_{Z^{\prime}}$, $g_{B-L}$), even under the most pessimistic conditions. The details of the analysis, simulation and results are presented below. 

\subsection{Analysis of the production mechanism and main background contributions}

\noindent Since this analysis closely follows the results of \cite{Queiroz:2024ipo}, the $Z^{\prime}$ production channel remains the Drell–Yan (DY) process. Consequently, the expected final state consists of a pair of highly energetic, oppositely charged reconstructed electrons or muons, i.e., $p \ p \rightarrow Z^{\prime} \rightarrow {l}^{+}{l}^{-}$ where ${l}^{\pm}={e}^{\pm},{\mu}^{\pm}$. For the cross-section calculations and the generation of kinematic and angular observables relevant for the analysis, Universal FeynRules Output (UFO) files containing the B-L model information were produced using the \textit{Feynrules Package} \cite{Degrande:2011ua, Alloul:2013bka}. These UFO files were subsequently used by the \textit{MadGraph5\_aMC@NLO} Monte Carlo Event Generator to simulate all the hard-scattering processes \cite{Alwall:2014hca}. The events were then processed by \textit{Pythia8} \cite{Sjostrand:2007gs} and \textit{Delphes3} \cite{deFavereau:2013fsa} for hadronization and incorporate detector effects\footnote{In the simulations, the NNPDF30\_nlo\_as\_0118 Parton Distribution Function (PDF) and the official Delphes card for the FCC-hh were used, ensuring that the Monte Carlo pseudo-events approximate as closely as possible the expected data at the FCC-hh with a center-of-mass energy of $\sqrt{s}$ = 100\ $\text{TeV}$.}. For each ($M_{Z'}$, $g_{B-L}$) scenario shown in Fig. \ref{cross_section_benchmarks_a}, $2 \times 10^5$ signal events were generated. On the other hand, a total of $4.6 \times 10^5$  background events were simulated, covering the main SM processes contributing to the same final state: irreducible DY, diboson ($WW$, $ZW$, $ZZ$), single-top quark ($tW$) and top quark pair ($t\bar{t}$) productions. 

Events were selected for analysis if they contained at least two high-energy electrons or two high-energy muons with opposite electric charge. Both signal and background events were initially generated using the \textit{Basic Selection} cut showed in Table \ref{selection_cuts_table_1}.

\begin{table}[h!]
\centering
\begin{tabular}{|c|c|c|c|c|}
\hline
\makecell{\textbf{Observable}} & \makecell{\textbf{Basic Sel.}} & \makecell{\textbf{Sel. 1}} & \makecell{\textbf{Sel. 2}} & \makecell{\textbf{Sel. 3}} \\
\hline
${p}_{T}(l^{\pm})>$ & 50 GeV & 100 GeV & 500 GeV & 1000 GeV \\
\hline
$|\eta(l^{\pm})|<$ & 6 & 5.5 & 5 & 4 \\
\hline
$\Delta R(l^{-},l^{+})>$ & 0.1 & 0.3 & 0.5 & 1 \\
\hline
\end{tabular}
\caption{Kinematic and angular cuts for the initial event selection analysis at FCC-hh.}
\label{selection_cuts_table_1}
\end{table}

\begin{figure*}[!ht]
    \centering
     \subfloat[]{\includegraphics[scale=0.45]{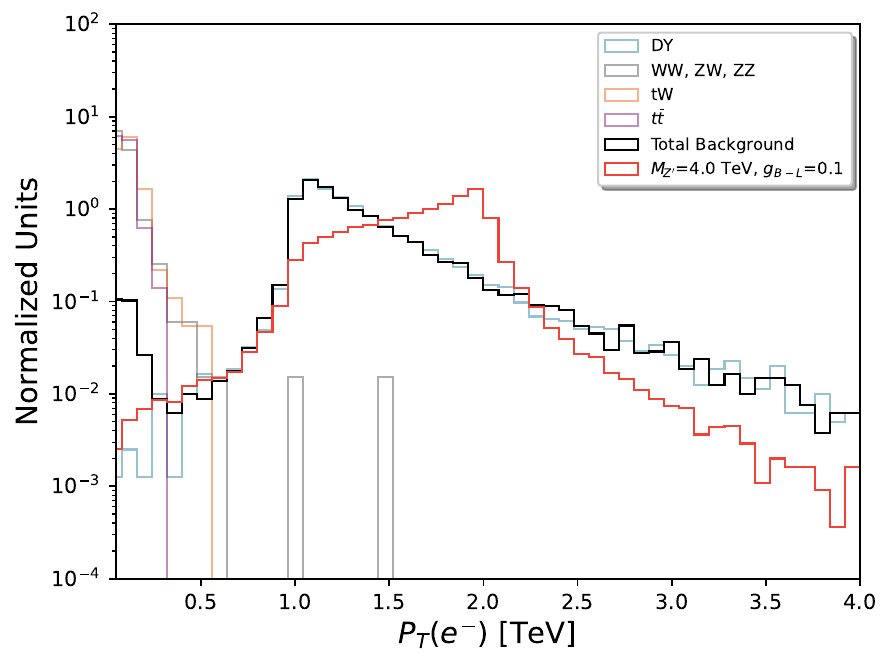}}
    \subfloat[]{\includegraphics[scale=0.45]{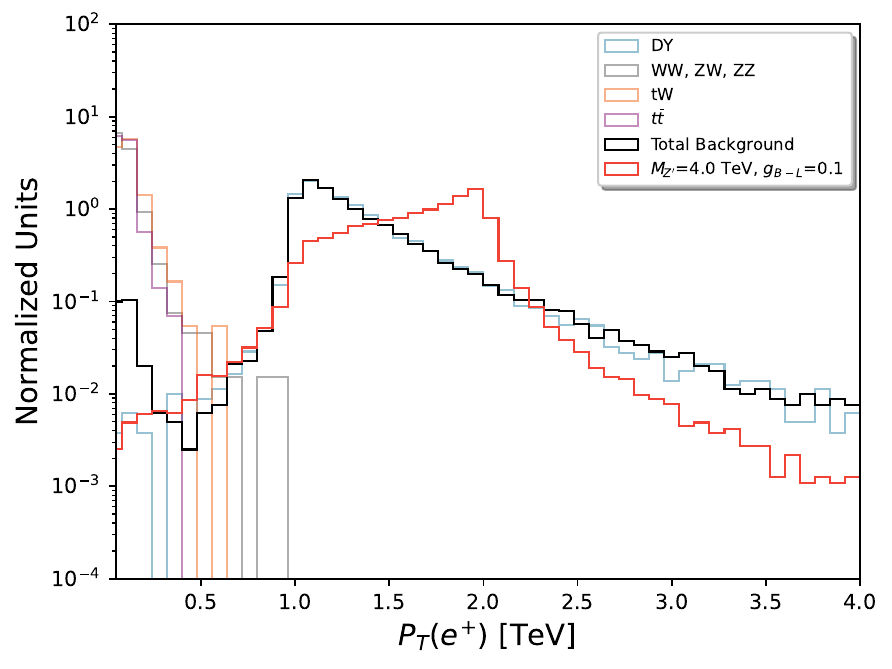}}
    \caption{Transverse momentum distribution $p_T$ of (a) electrons and (b) positrons at FCC-hh, showing the different background contributions and the (4\ TeV, 0.1) scenario.}
\label{pt_electron_positron_with_basic_cut}
\end{figure*}

\begin{figure*}[!ht]
    \centering
    \subfloat[]{\includegraphics[scale=0.45]{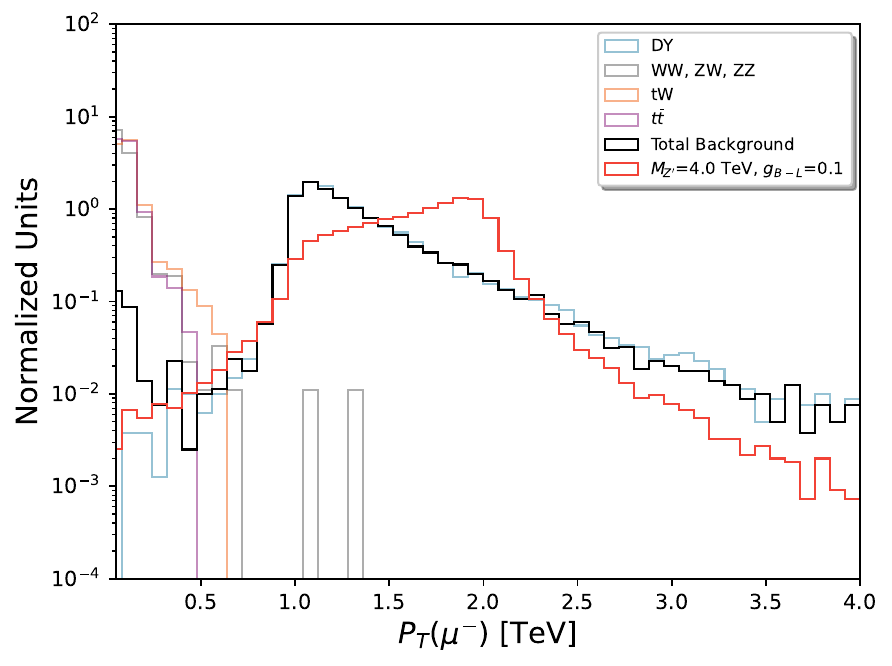}}
    \subfloat[]{\includegraphics[scale=0.45]{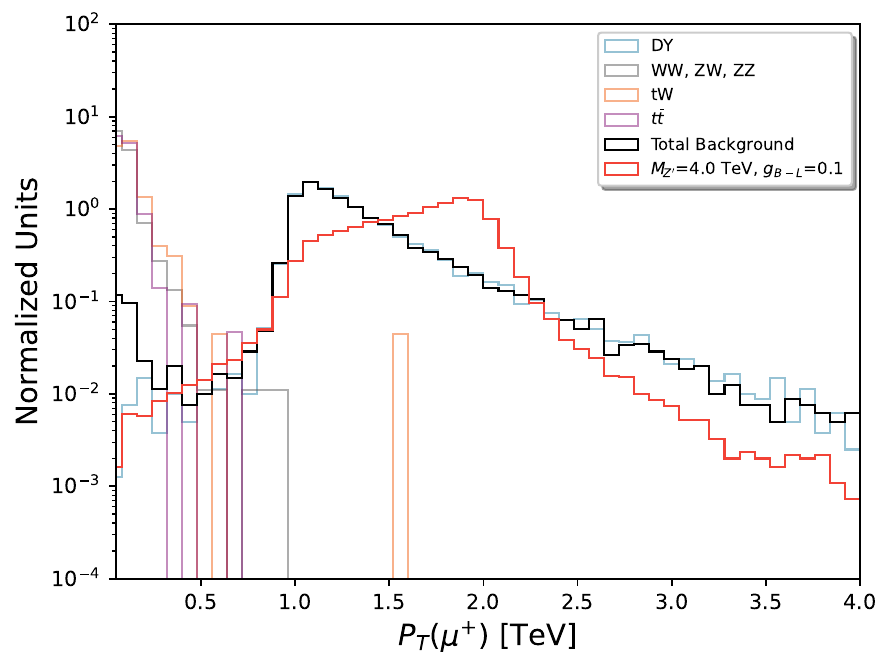}}
    \caption{Transverse momentum distribution $p_T$ of (a) muons and (b) antimuons at FCC-hh, showing the different background contributions and the (4\ TeV, 0.1) scenario.}
\label{pt_muon_antimuon_with_basic_cut}
\end{figure*}

\begin{figure}[!ht]
    \centering
    \includegraphics[scale=0.45]{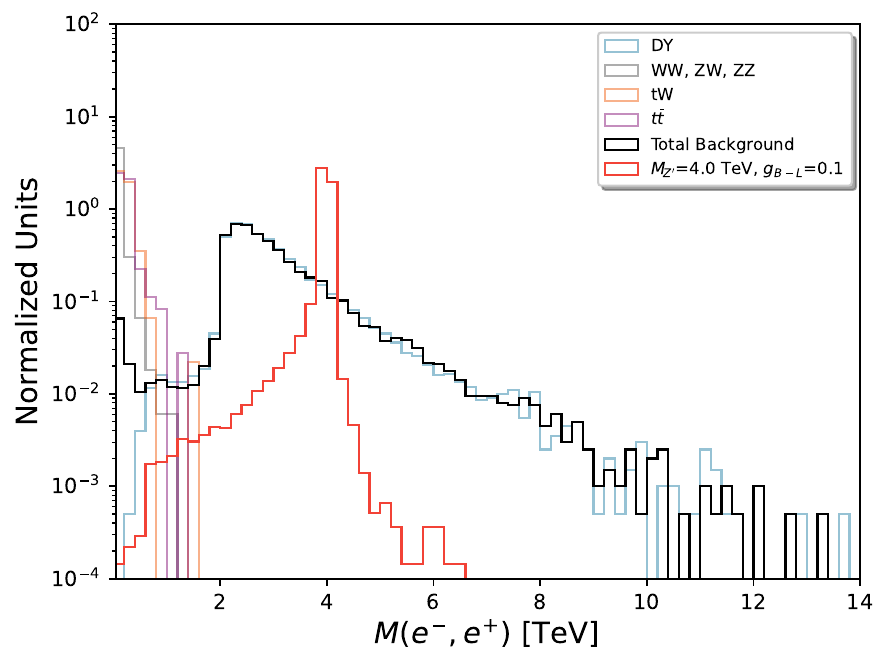}
    \caption{Invariant mass distribution of electron-positron pairs at FCC-hh, showing different background contributions and the (4\ TeV, 0.1) scenario.}
    \label{invmass_with_basic_cut_ee}
\end{figure}

\begin{figure}[!ht]
    \centering
    \includegraphics[scale=0.45]{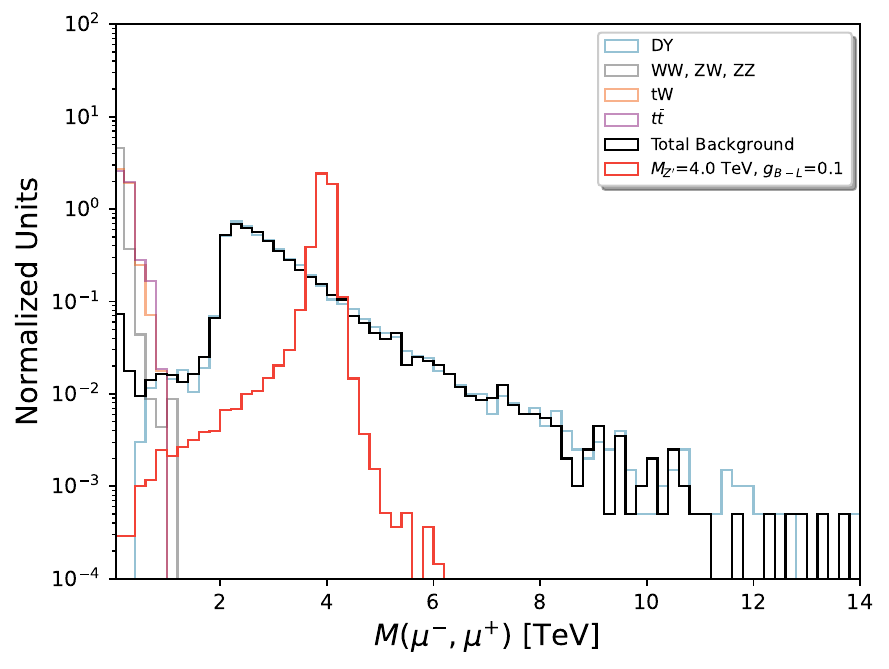}
    \caption{Invariant mass distribution of muon-antimuon pairs at FCC-hh, showing different background contributions and the (4\ TeV, 0.1) scenario.}
    \label{invmass_with_basic_cut_mumu}
\end{figure}

\begin{figure*}[!ht]
    \centering
    \subfloat[]{\includegraphics[scale=0.45]{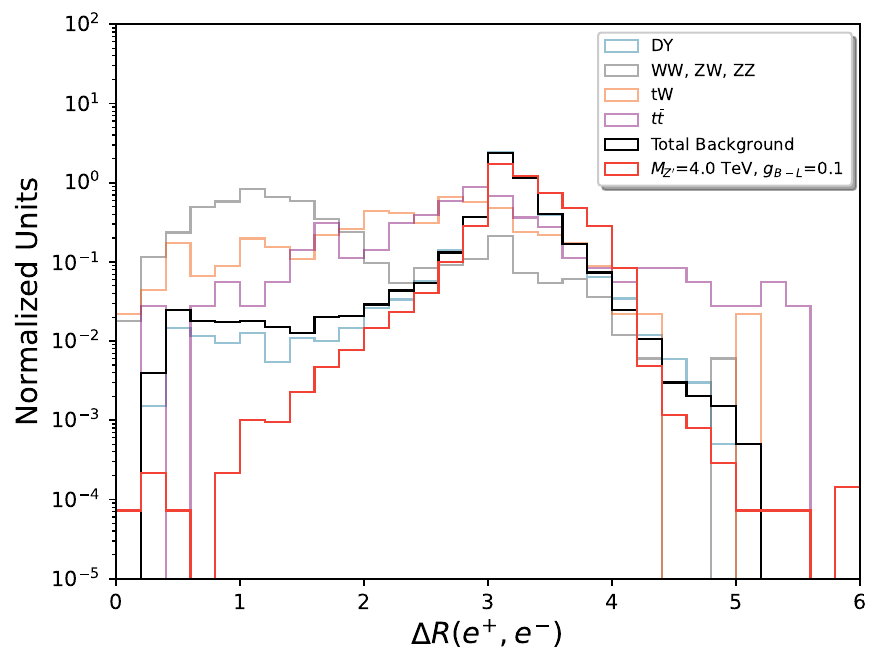}}
    \subfloat[]{\includegraphics[scale=0.45]{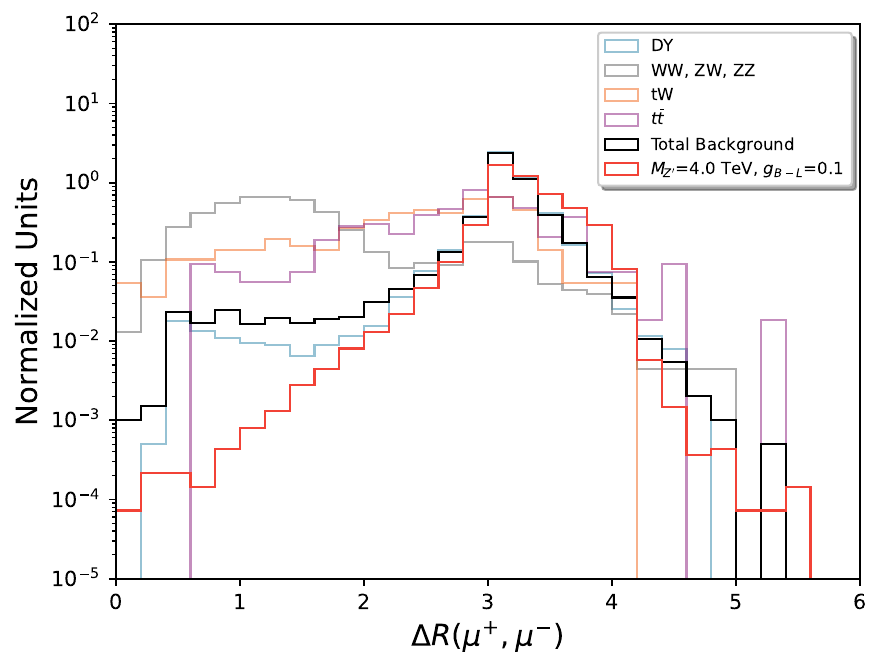}}
    \caption{Separation in the $\eta \times \phi$ plane, defined by $\sqrt{{\Delta\phi}^2 + {\Delta\eta}^2}$, between (a) electron-positron and (b) muon-antimuon pairs at FCC-hh, showing different background contributions and the (4\ TeV, 0.1) scenario.}
\label{deltar_with_basic_cut}
\end{figure*}

\begin{figure*}[!ht]
    \centering
    \subfloat[]{\includegraphics[scale=0.45]{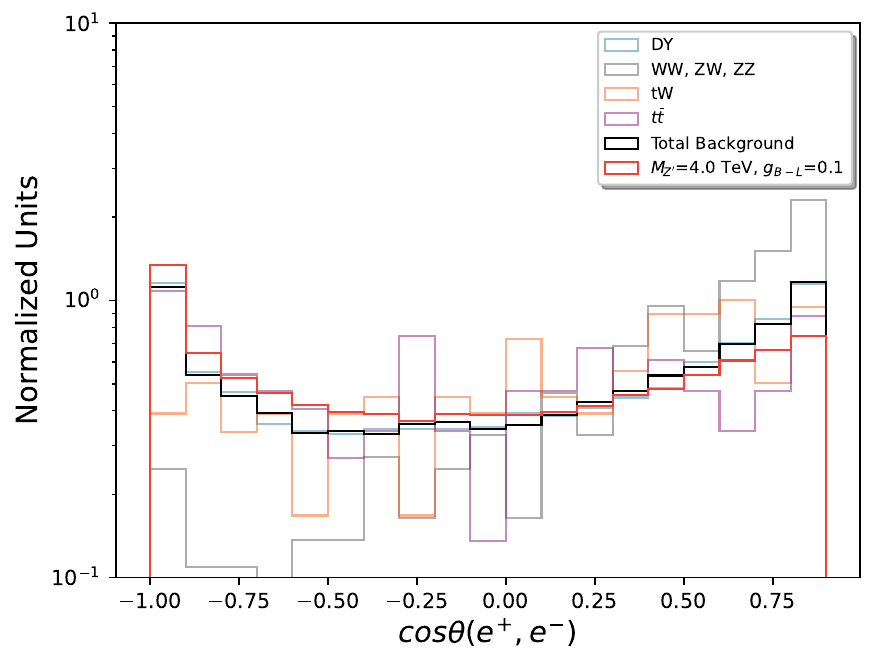}}
    \subfloat[]{\includegraphics[scale=0.45]{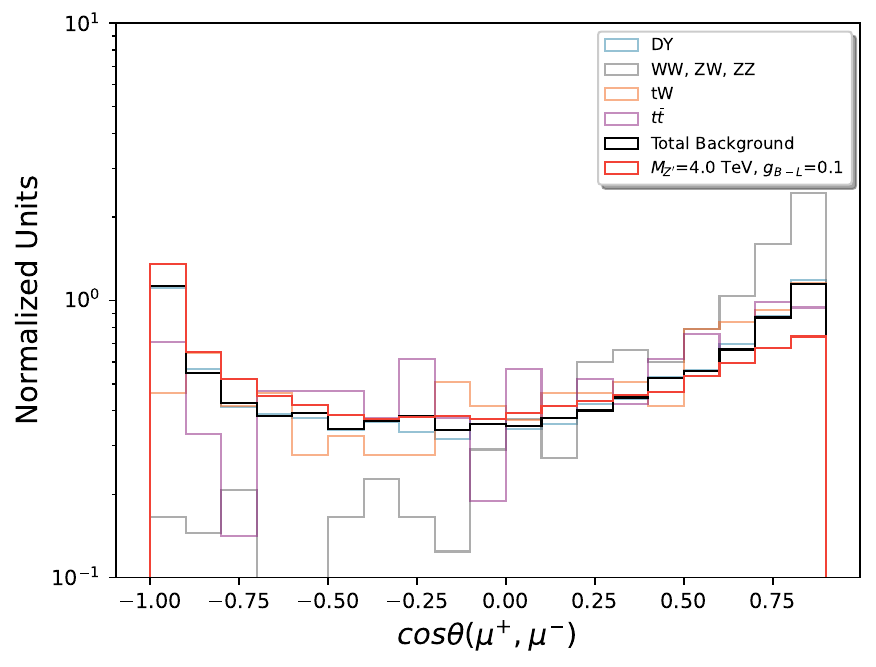}}
    \caption{Cosine of angular separation between (a) electron-positron and (b) muon-antimuon pairs at FCC-hh, showing different background contributions and the (4\ TeV, 0.1) scenario.}
\label{costheta_with_basic_cut}
\end{figure*}

\noindent Figures \ref{pt_electron_positron_with_basic_cut}–\ref{costheta_with_basic_cut} show the transverse momentum distributions $p_T$ for electrons and muons, as well as the invariant mass $M$, the $\Delta {R}$ separation in the $\eta \times \phi$ plane, and cosine of the separation angle, $\cos{\theta}$, between ${e}^{+}{e}^{-}$ and ${\mu}^{+}{\mu}^{-}$ pairs. In all plots, the different background contributions are compared with the (4\ TeV, 0.1) scenario. 

\noindent The transverse momentum distributions for the signal, shown in Figs. \ref{pt_electron_positron_with_basic_cut} and \ref{pt_muon_antimuon_with_basic_cut}, exhibits a Jacobian peak centered at $p_T({l}^{\pm})\sim M_{{Z}^{\prime}}/2 = 2$ TeV, as expected for this type of kinematics. In contrast, the $p_T$ distributions for backgrounds suggest that a hard cut on $p_T({l}^{\pm})$ would effectively suppress contributions from top quark pair, single-top quark, and diboson processes, leaving DY production as the dominant irreducible background.

\noindent The invariant mass distributions $M$ for ${e}^{+}{e}^{-}$ and ${\mu}^{+}{\mu}^{-}$ pairs, shown in Figs. \ref{invmass_with_basic_cut_ee} and \ref{invmass_with_basic_cut_mumu}, exhibit a resonance centered at $M_{{Z}^{\prime}} \sim 4$ TeV. The signal distribution shows a tail extending toward lower invariant masses, consistent with energy losses due to final-state radiation (FSR). In the high-mass regime, the dominant background contribution is from DY processes. 

\noindent On the other hand, the angular distributions, $\Delta R$ and $\cos \theta$, between $e^+e^-$ and $\mu^+\mu^-$ pairs showed in Figs. \ref{deltar_with_basic_cut} and \ref{costheta_with_basic_cut}, respectively, exhibit the expected back-to-back kinematics. Both signal and irreducible DY background contribution show a pronounced peak around of $\Delta R \sim \pi$, a consequence of $|\Delta \phi|$ dominating in $|\Delta \phi| \sim \pi$. This behavior arises because leptons produced here tend to be emitted in opposite directions in the transverse plane. The remaining background contributions ( $t\bar{t}$, $tW$, $WW$, $WZ$, $ZZ$) exhibit much broader and flatter shapes. The $\cos \theta$ distribution dominating in $\cos \theta \sim -1$ for signal and DY background contribution follow the same back-to-back kinematics. The $t\bar{t}$ and $tW$ background contributions shows a more dispersed distribution and is less angularly correlated. In contrast, diboson background contribution shows a distinct kinematic behavior, characterized by an excess in the region $\cos\theta > 0.1$. This behavior is due to the fact that the bosons produced here are boosted, especially from ZZ contributions, causing the leptons to tend to be emitted collinearly. 

\noindent To suppress background sources that could obscure potential ${Z}^{\prime}$ signals at the FCC-hh, the remaining selection cuts described in Table \ref{selection_cuts_table_1} (\textit{Selection 1, 2} and \textit{3}) were applied. The cross-section values obtained after these initial selection cuts are presented in Table \ref{selection_cuts_table_2} for three representative ($M_{Z^{\prime}}$, $g_{B-L}$) scenarios and the different background contributions. \newline

\begin{table}[h]
\centering
\begin{tabular}{|c|c|c|c|c|}
\hline
\makecell{\textbf{Process}} & \makecell{\textbf{Basic Sel.}} & \makecell{\textbf{Sel. 1}} & \makecell{\textbf{Sel. 2}} & \makecell{\textbf{Sel. 3}} \\
\hline
(4TeV,\ 0.1) & 1.13$\times 10^{-2}$ & 1.13$\times 10^{-2}$ & 1.13$\times 10^{-2}$ & 1.07$\times 10^{-2}$ \\
\hline
(8TeV,\ 0.4) & 1.36$\times 10^{-2}$ & 1.35$\times 10^{-2}$ & 1.35$\times 10^{-2}$ & 1.33$\times 10^{-2}$ \\
\hline
(12TeV,\ 0.3) & 1.22$\times 10^{-3}$ & 1.22$\times 10^{-3}$ & 1.21$\times 10^{-3}$ & 1.20$\times 10^{-3}$ \\
\hline
$DY$ & 5.63$\times 10^{-3}$ & 5.63$\times 10^{-3}$ & 5.59$\times 10^{-3}$ & 4.93$\times 10^{-3}$ \\
\hline
$WW$ & 1.93 & 0.15 & 0 & 0 \\
\hline
$ZW$ & 1.68 & 0.21 & 0 & 0 \\
\hline
$ZZ$ & 1.10 & 0.14 & 0 & 0 \\
\hline
$tW$ & 11.18 & 2.29 & 0 & 0 \\
\hline
$ t\bar{t}$ & 86.17 & 12.91 & 0 & 0 \\
\hline
\end{tabular}
\caption{Cross-sections, in pb, for three representative (${M}_{{Z}^{\prime}}$, ${g}_{B-L}$) signal scenarios and background processes after successive selection cuts shown in Table \ref{selection_cuts_table_1}. }
\label{selection_cuts_table_2}
\end{table}

\begin{figure}[!ht]
    \centering
    \includegraphics[scale=0.45]{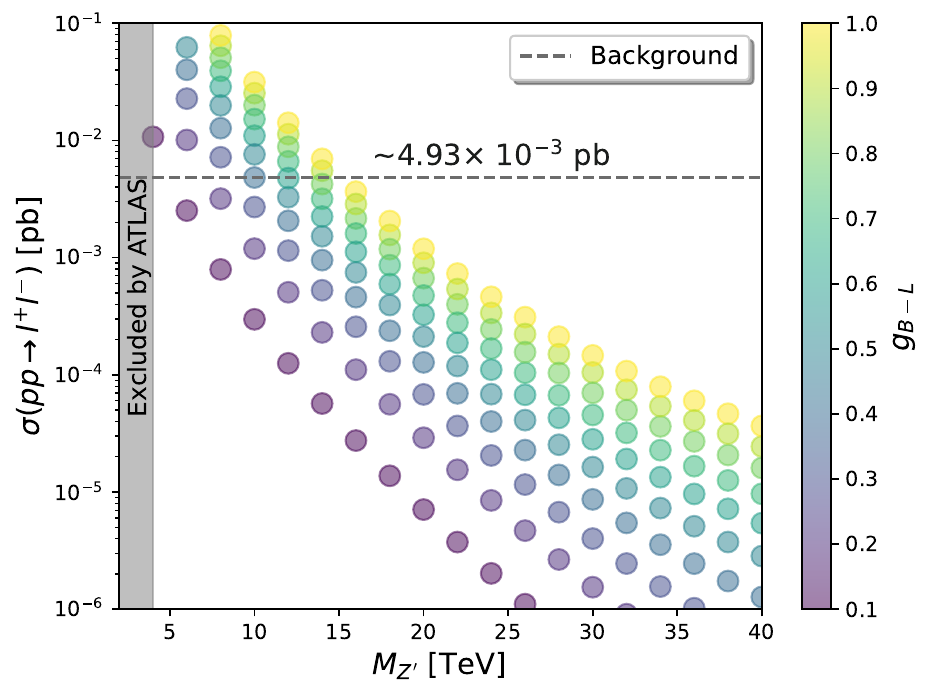}
    \caption{Production cross-section for different (${M}_{{Z}^{\prime}}$, ${g}_{B-L}$) scenarios after \textit{Sel. 3}. The black dashed line is the irreducible SM contribution.}
    \label{cross_section_benchmarks_b}
\end{figure}

\noindent With this initial cut-flow we observe that \textit{Selection 3} effectively suppresses most background contributions, reducing the total background cross-section to $4.93\times{10}^{-3}$ pb. The cross-section values for several ($M_{Z'}$, $g_{B-L}$) scenarios, after \textit{Sel. 3}, are shown in Fig. \ref{cross_section_benchmarks_b}.

\subsection{Details of the Analysis}

\begin{figure*}[!ht]
     \centering
      \subfloat[]{\includegraphics[scale=0.45]{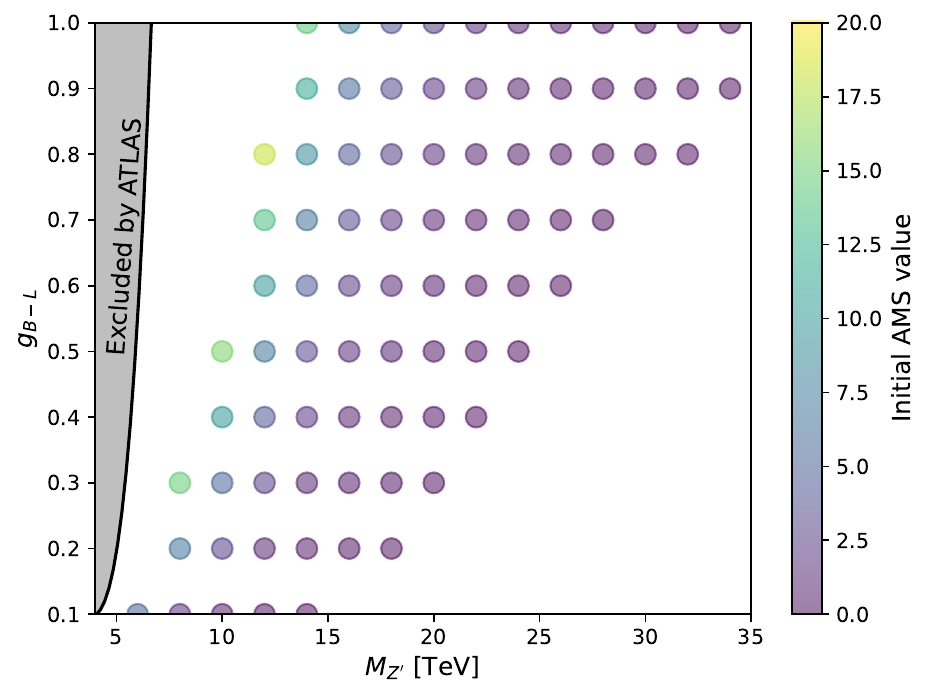}
      \label{select_benchmarks_after_ams_1}}
     \subfloat[]{\includegraphics[scale=0.45]{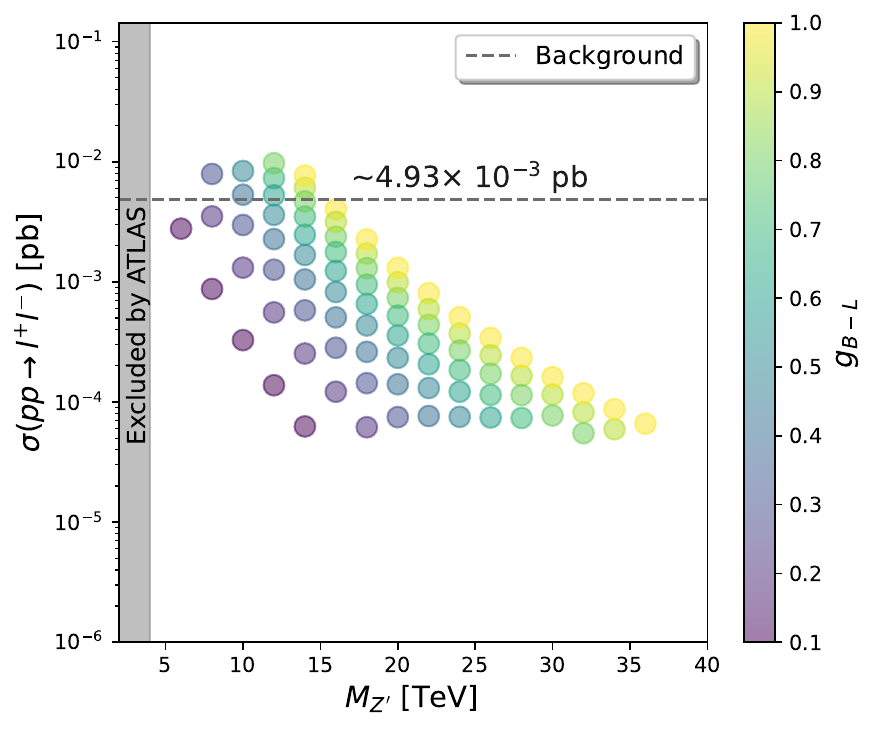}
      \label{select_benchmarks_after_ams_2}}
     \caption{(a) All ($M_{Z'}$, $g_{B-L}$) scenarios that fall within the initial sensitivity band ($0.1\sigma < AMS < 20\sigma$) and (b) cross-section values corresponding to these points.
     }
     \label{select_benchmarks_after_ams_12}
\end{figure*}

\noindent Assuming that FCC-hh operates at an integrated luminosity of $\mathcal{L}_{int} = $ 3~$ab^{-1}$ and a center-of-mass energy of $\sqrt{s}=100$, not all scenarios in Fig. \ref{cross_section_benchmarks_b} are experimentally accessible. To estimate the projected sensitivity, we will initially consider a certain \textit{sensitivity band} where the FCC-hh would observe signals of new physics between 0.1$\sigma$ and 20$\sigma$ of statistical significance after \textit{Sel. 3}. Figures \ref{select_benchmarks_after_ams_1} and \ref{select_benchmarks_after_ams_2} show the scenarios that fall within this band and their respective cross-section values.

\noindent Here, the significance is calculated using,

\begin{equation}
     AMS = \frac{\mathcal{L}_{int} \times {\epsilon}_{S} \times {\sigma}_{S}}{\sqrt{\mathcal{L}_{int} \times {\epsilon}_{B} \times {\sigma}_{B}+{({\epsilon}_{B}^{sys} \times \mathcal{L}_{int} \times {\epsilon}_{B} \times {\sigma}_{B})}^{2}}},
    \label{significance}
\end{equation}

\noindent where ${\epsilon}_{B}^{sys}$ denotes the background systematic error\footnote{Although very precise measurements are expected for the FCC-hh, in this work we assume background systematic error of 10\%, to be conservative}, $\mathcal{L}_{int}$ is the integrated luminosity, and $\sigma_B$ ($\sigma_S$) and $\epsilon_B$ ($\epsilon_S$) represent the background (signal) cross-section and selection cut efficiency, respectively. At this point, it can be noted that the FCC-hh can probe signals of new physics with more than 20$\sigma$, considering only \textit{Sel. 3}. It is essential to note that these discovery scenarios, which demonstrate high statistical significance after \textit{Selection} 3 are amenable to PDF uncertainties. In contrast, the most interesting scenarios to analyze here are those that have cross-section values smaller than the background one (or scenarios with very low statistical significance) according to Fig. \ref{select_benchmarks_after_ams_12}, in order to demonstrate the FCC-hh potential to exclude with 95\% C.L. or discovery with 5$\sigma$ this type of scenario\footnote{This is because, as we will demonstrate, with an optimized selection cut strategy, we can bring many of these scenarios to 2$\sigma$ and 5$\sigma$ of statistical significance, even with a luminosity lower than expected integrate luminosity.}.

\begin{figure}[!ht]
    \centering
    \includegraphics[scale=0.45]{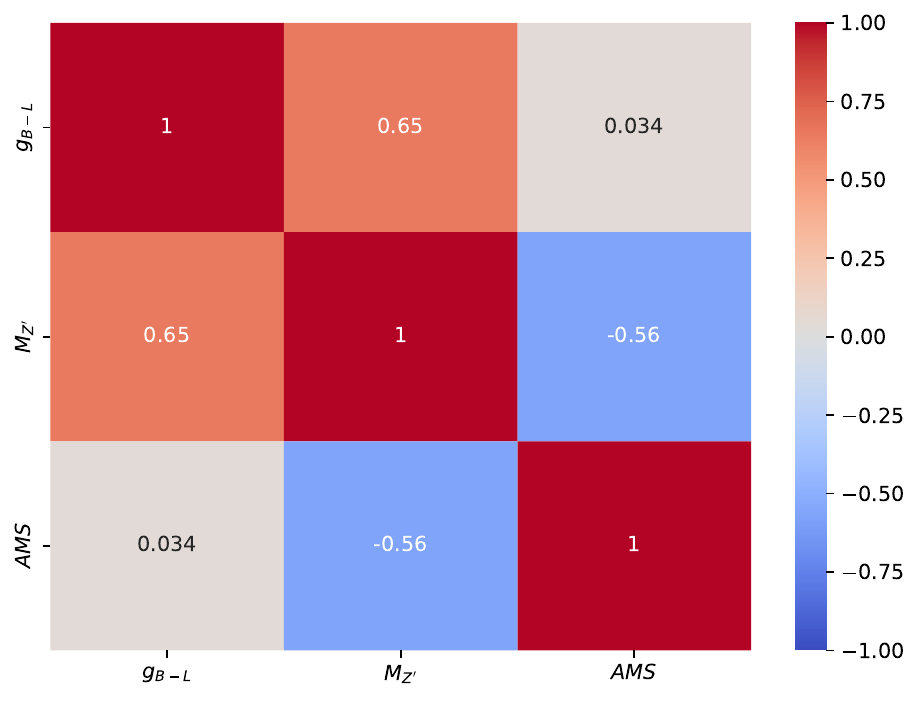}
    \caption{Correlation matrix between the parameters ($M_{Z'}$, $g_{B-L}$) and the statistical significance calculated after Selection 3.}
    \label{correlation_sigma_mass_coupling_after_ams}
\end{figure}

\noindent On the other hand, we can obtain information about the kinematics of the final leptons through the correlation between $AMS$, the mass ${M}_{{Z}^{\prime}}$ and coupling $g_{B-L}$. Fig. \ref{correlation_sigma_mass_coupling_after_ams} show a strong negative correlation between the mass ${M}_{{Z}^{\prime}}$ and $AMS$. This relation arises from the parton distribution functions: producing a heavy ${Z}^{\prime}$ vector boson in the FCC-hh at 100 TeV requires quarks within the proton to carry a large momentum fraction, causing the production cross-sections to drop considerably with $M_{Z^{\prime}}$, and consequently reducing the statistical significance. On the other hand, the weak correlation between $AMS$ and $g_{B-L}$ suggests that, within the sensitivity region considered, $g_{B-L}$ is completely correlated with ${M}_{{Z}^{\prime}}$ to compensate the PDF suppression. This result indicates that the observability at the FCC-hh of a $Z^{\prime}$ vector boson from the $B-L$ gauge symmetry depends predominantly on its mass.

\noindent In order to maximize the sensitivity to a new ${Z}^{\prime}$ from the B-L model at the FCC-hh, a random search algorithm was implemented. For each one of the ($M_{Z^{\prime}}$, $g_{B-L}$) scenarios, the algorithm performed $4 \times 10^5$ random searches that identified the best set of kinematic and angular cuts that maximize signal efficiency; that is, it finds the cuts that best capture events from the production of a $Z^{\prime}$, while minimizing background ome. The results of this search for differente representative scenarios are shown in Table \ref{best_cuts_representative_scenarios}, where we observe the regions of the parameter space identified by the algorithm that favor the observation of new physics. 

\begin{table*}[ht]
\centering
\begin{tabular}{|c|c|c|c|c|c|c|c|c|c|}
\hline
$M_{Z^{\prime}}$ & $g_{B-L}$ & ${p}_{T}({l}^{\pm})>$ & $cos\theta({l}^{+},{l}^{-})$ & $M({l}^{+},{l}^{-})$ & $\Delta R({l}^{+},{l}^{-})$ & $N_S$ & ${\epsilon}_S$ [\%] & $N_B$ & ${\epsilon}_B$ [\%] \\ \hline
4000 & 0.1 & 1694 & 0.76$\pm$0.25 & 4039$\pm$201 & 1.44$\pm$1.91 & 4750 & 14.8 & 132 & 0.89 \\ \hline
8000 & 0.3 & 3347 & -0.71$\pm$0.29 & 8080$\pm$401 & 4.98$\pm$1.74 & 301 & 3 & 31 & 0.20 \\ \hline
12000 & 0.9 & 4781 & -0.76$\pm$0.28 & 13021$\pm$1162 & 1.72$\pm$1.47 & 718 & 4.5 & 2 & 0.013 \\ \hline
20000 & 0.3 & 7910 & -0.70$\pm$0.29 & 19700$\pm$2991 & 2.94$\pm$1.32 & 30 & 31.7 & 1 & 0.006  \\ \hline
32000 & 0.8 & 10010 & -0.70$\pm$0.29  & 29409$\pm$4799 & 3.76$\pm$1.43 & 11 & 16.2 & 1 & 0.005 \\ \hline
\end{tabular}
\caption{Best kinematic and angular cuts for representative ($M_{Z^{\prime}}$, $g_{B-L}$) scenarios. The transverse momentum and invariant mass are given in GeV.}
\label{best_cuts_representative_scenarios}
\end{table*}

\noindent The Table \ref{best_cuts_representative_scenarios} shows the optimized kinematic and angular selection cuts that produce the best statistical significance after sequential application of all cuts, where $N_S$ and $N_B$ are the number of signal and background events expected to be observed in the FCC-hh. We can observe that the selection cut efficiencies of the signal ${\epsilon}_S$ vary from 3\% to 31.7\% for the scenarios shown, while the background cut efficiencies are always suppressed, demonstrating the effectiveness of the cuts.

\begin{figure*}[!ht]
    \centering
     \subfloat[]{\includegraphics[scale=0.45]{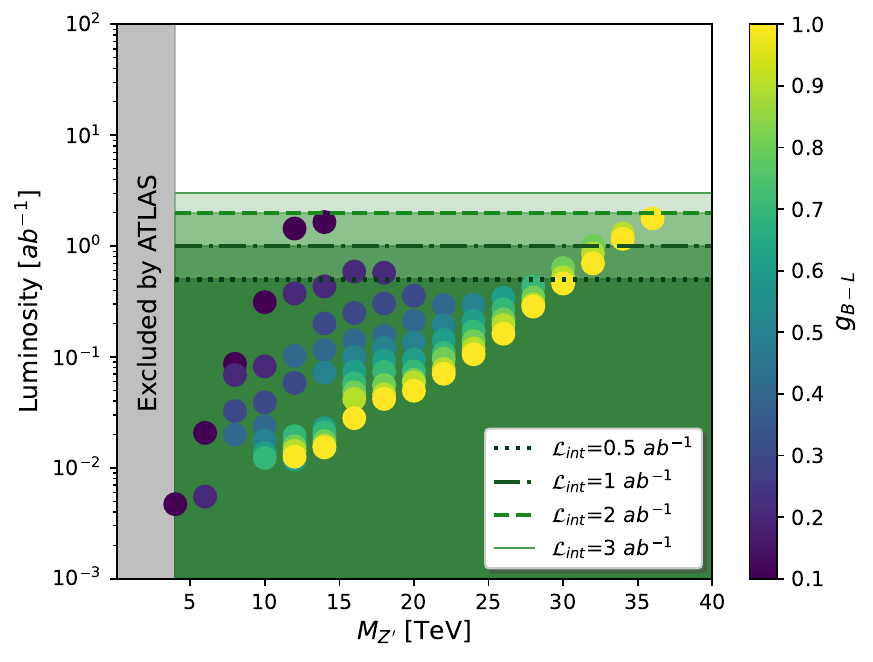}
     \label{select_benchmarks_luminosity_2sigma}}
    \subfloat[]{\includegraphics[scale=0.45]{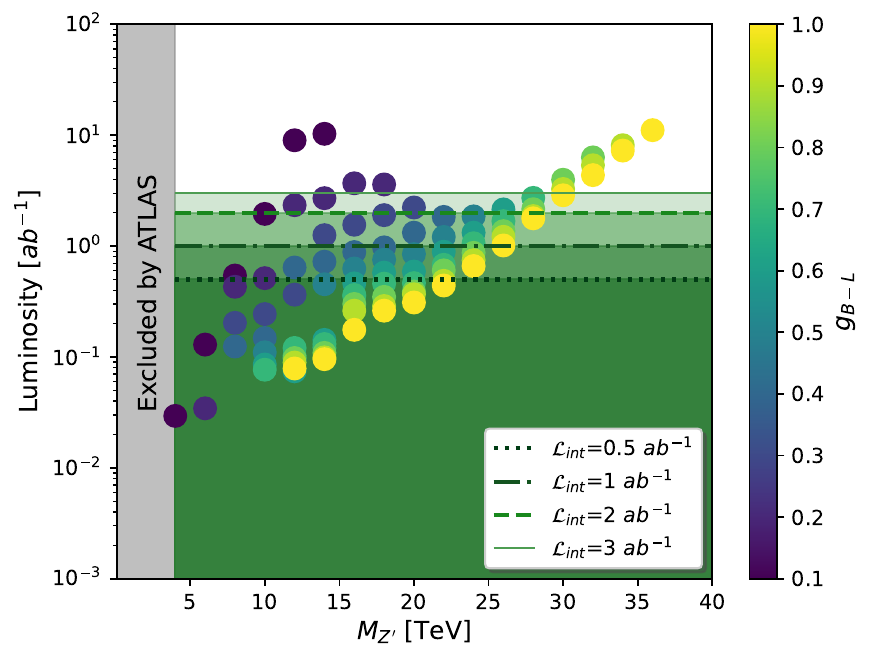}
     \label{select_benchmarks_luminosity_5sigma}}
    \caption{The required luminosity for (a) exclusion at 95\% C.L. and (b) discovery with 5$\sigma$ of a new $Z^{\prime}$ from the B-L model at the FCC-hh.}
    \label{select_benchmarks_luminosity_2sigma_5sigma}
\end{figure*}

\noindent Finally, using the results obtained by the random search algorithm, in Fig. \ref{select_benchmarks_luminosity_2sigma_5sigma} we estimate the required luminosity in the FCC-hh ($\sqrt{s} = 100$ TeV) to probe the parameter space ($M_{Z^{\prime}}$, $g_{B-L}$) for two different statistical scenarios: exclusion with 95\% of C.L. (Fig. \ref{select_benchmarks_luminosity_2sigma}) or discovery potential of observing signals with 5$\sigma$ of statistical significance (Fig. \ref{select_benchmarks_luminosity_5sigma}).


\noindent The green lines represent different stages of FCC-hh operation, varying from $0.5$~$ab^{-1}$ to $3$~$ab^{-1}$. In Fig. \ref{select_benchmarks_luminosity_2sigma_5sigma} we can observe that with an integrated luminosity of $\mathcal{L}_{int} = $3~$ab^{-1}$ the FCC-hh is able to exclude $Z^{\prime}$ with masses up to $M_{Z^{\prime}}\sim$ 40 TeV for strong couplings $g_{B-L} \sim 1$ as well as exclude masses up to $M_{Z^{\prime}}\sim$ 15 TeV for weak couplings $g_{B-L} \sim 0.1$. On the other hand, within a discovery scenario at 5$\sigma$ (Fig. \ref{select_benchmarks_luminosity_5sigma}), being statistically more restrictive, the FCC-hh reduces its mass reach and would achieve $Z^{\prime}$ scenarios on the order of $M_{Z^{\prime}}\sim$ 30 TeV for couplings $g_{B-L} \sim 0.8$ and weak coupling scenarios $g_{B-L} \sim 0.1$ of up to $M_{Z^{\prime}}\sim$ 10 TeV. This result demonstrates that, with only three years of operation at an integrated luminosity of $\mathcal{L}_{int} = 3$~$ab^{-1}$, the FCC-hh can not only exclude $Z^{\prime}$ up to 40 TeV at 95\% C.L., but also probe scenarios up to 30 TeV with 5$\sigma$. 

\section{Conclusions}
\label{conclusions}

\noindent We have performed a phenomenological analysis of the FCC-hh ($\sqrt{s}$ = 100 TeV) reach to a new heavy neutral vector boson, $Z^{\prime}$, predicted by the $U(1)_{B-L}$ model. Focusing on the parameter space currently not excluded by LHC data and assuming the expected first years of operation for the FCC-hh ($\mathcal{L}_{int} = 3$~$ab^{-1}$) our results demonstrate that FCC-hh can exclude a large part of the viable parameter space of the model. 

\noindent To this end, we analyzed the dilepton channel and performed full simulations for different ($M_{Z^{\prime}}$, $g_{B-L}$) scenarios and different background sources that contribute to the same final state. We observed that under a set of initial selections cuts, we were able to reduce a large part of the background contributions, while maintaining the DY contributions as dominant. Our results show that with such integrated luminosity FCC-hh can exclude $Z^{\prime}$ masses up to $\sim$ 40 TeV for strong couplings $g_{B-L}$ $\sim$ 1 at 95\% C.L.. Concerning the discovery potential, FCC-hh can achieve 5$\sigma$ signals for masses up to $\sim$ 30 TeV for $g_{B-L}$ $\sim$ 0.8, up to $\sim$ 15 TeV for $g_{B-L}$ $\sim 0.1$.

\noindent Clearly, FCC-hh is projected as an indispensable tool for the future of particle physics with unprecedented power to probe physics beyond the standard model such as those predicted by the $U(1)_{B-L}$ gauge symmetry.


\section*{Acknowledgements}

\noindent FSQ thanks CERN and the Max Planck Institute for Nuclear Physics for the hospitality where this project was partly conducted. The authors thank Alfonso Zerwekh and Patricia Teles for discussions. The work of Y.M. Oviedo-Torres, FSQ and J. Zamora-Saa was funded by ANID - Millennium Science Initiative Program - ICN2019 044. J. Zamora-Saa was partially supported by FONDECYT grant 1240216 and 1240066. Ricardo C. Silva was supported by a CAPES Grant (88887.966347/2024-00). Y.M Oviedo-Torres was supported by FONDECYT grant No. 3250068. FSQ is supported the Simons Foundation (Award Number: 1023171-RC) and IIF-FINEP grant 213/2024. FSQ acknowledges FAPESP Grants 2018/25225-9, 2021/01089-1, 2023/01197-4, ICTP-SAIFR FAPESP Grants 2021/14335-0, CNPQ Grants 403521/2024-6, 408295/2021-0, 403521/2024-6, 406919/2025-9, 351851/2025-9. 

\newpage

\bibliography{referencias}

\end{document}